\documentclass[final,5p,times,twocolumn]{elsarticle}
\usepackage{amssymb}
\usepackage{color}
\usepackage{amsmath}
\usepackage{multirow}
\usepackage{booktabs}
\usepackage{braket,ulem}

\usepackage{graphicx}
\usepackage[pdfstartview=FitH, colorlinks,
 linkcolor={red!60!black!},
 anchorcolor={green!80!black!},
 urlcolor={blue!80!black!},
 citecolor={blue!50!black!}]{hyperref}
\usepackage[table]{xcolor}
\journal{Science Bulletin}

\usepackage{lineno}
\usepackage{color} 
\biboptions{numbers,sort&compress}

\begin{document}
\begin{frontmatter}

\title{Remarks on strong phase shifts in weak nonleptonic baryon decays}

\author[1,2]{Hong-Jian Wang}
\author[1,2]{Pei-Rong Li\corref{cor1}}
\author[3]{Xiao-Rui Lyu\corref{cor2}}
\author[4]{Jusak Tandean}
\author[3,4]{Hai-Bo Li}

\address[1]{\smash{School of Nuclear Science and Technology, Lanzhou University, Lanzhou 730000, China}}
\address[2]{\smash{MOE Frontiers Science Center for Rare Isotopes, Lanzhou University, Lanzhou 730000, China}}
\address[3]{\smash{School of Physical Sciences, University of Chinese Academy of Sciences, Beijing 100049, China}}
\address[4]{\smash{Institute of High Energy Physics, Chinese Academy of Sciences, Beijing 100049, China}}
\cortext[cor1]{Email: prli@lzu.edu.cn.}
\cortext[cor2]{Email: xiaorui@ucas.ac.cn.}

\end{frontmatter} 


The matter-antimatter asymmetry observed in nature is one of the biggest mysteries in physics, and the resolution of this puzzle could be one of the necessary paths to identifying the ultimate theory of the universe. The violation of the combined charge-conjugation and parity~($CP$) symmetry is one of the required ingredients for explaining this asymmetry~\cite{Sakharov:1967dj}. Although $CP$ violation has been discovered in meson processes~\cite{KTeV:1999kad,LHCb:2019hro,LHCb:2019jta}, so far only evidence for $CP$ violation in the baryon decays $\Lambda_b^0\to p\pi^-\pi^+\pi^-$ and $\Lambda_b^0\to\Lambda K^+K^-$ has been found~\cite{LHCb:2016yco,LHCb:2024yzj}.

With the development of experimental technology in recent years, data samples of baryons have become increasingly large, and polarization studies of baryon weak decays with sufficient precision have emerged (e.g., \cite{Schonning:2023mge}). Thanks to the discovery of parity violation in weak decays, the construction of $CP$-observables through polarization parameters has become an important way to search for the $CP$-violation phenomenon among baryons.

\begin{figure}[ht] \centering \vspace{-1ex} 
\includegraphics[trim=0mm 1mm 0mm 4pt,clip,width=0.9\linewidth]{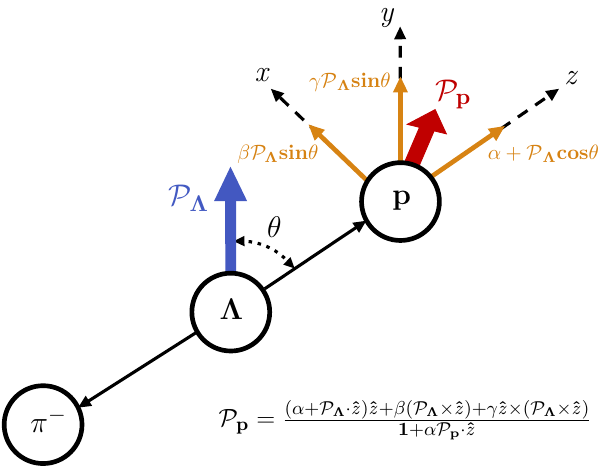} \vspace{-5pt}
\caption{Generation and transmission of polarization in weak decay.} \label{polarization}
\end{figure}

In 1957 Lee and Yang did the first general partial-wave analysis on a spin-1/2 hyperon decaying into a nucleon and pion~\cite{Lee:1957qs}. 
This is illustrated in Fig.\,\ref{polarization} for $\Lambda\to p\pi^-$, but applies more generally to any weak decay involving hadrons of the same respective spins and parities, $1/2^+\to1/2^++0^-$.
As the figure shows, the $\Lambda$ and proton polarizations, ${\cal P}_\Lambda$ and ${\cal P}_p$, are related in terms of the parameters~\cite{Lee:1957qs,ParticleDataGroup:2024cfk} 
\begin{equation} \label{Polarization}
\alpha = \frac{2\,\text{Re}(S^{*}P)}{|S|^{2}+|P|^{2}} \,, ~~~~ \beta = \frac{2\,\text{Im}(S^{*}P)}{|S|^{2}+|P|^{2}} \,, ~~~~ \gamma = \frac{|S|^{2}-|P|^{2}}{|S|^{2}+|P|^{2}} 
\end{equation}
satisfying \,$\alpha^2+\beta^2+\gamma^2=1$, where $S$ and $P$ are the {\tt S}- and {\tt P}-wave amplitudes for the decay corresponding to the final hadrons' relative orbital angular momenta of 0 and~1, respectively.  

It is quite common to express $S$ and $P$ as~\cite{ParticleDataGroup:2024cfk}
\begin{align} \label{DefineI}
S_1^{} & \,=\, |S|\, e^{i\delta_S} \,, & P_1^{} & \,=\, |P|\, e^{i\delta_P} \,, & 
\end{align}
where, if $CP$ is conserved, $\delta_{S,P}$  are phase angles arising from final-state strong interactions.
The subscript `1' indicates the first parameterization form for $S$ and $P$, with the polarization parameters similarly subscripted. 
The difference  \,$\delta_P-\delta_S$\, is termed the phase shift. 
Accordingly, $\alpha$ and $\beta$ in Eq.\,(\ref{Polarization}) become
\begin{equation} \label{DefineIdirive}
\alpha_1^{} = \frac{2|S||P| \cos(\delta_P-\delta_S)}{|S|^{2}+|P|^{2}} \,, ~~~~ \beta_1^{} = \frac{2|S||P| \sin(\delta_P-\delta_S)}{|S|^{2}+|P|^{2}} \,.
\end{equation}
From their ratio, one arrives at \,$\delta_P-\delta_S=\text{arctan}(\beta_1/\alpha_1)+n\pi$, where \,$n=0,\pm1,\pm2,\raisebox{0.5ex}{...}$, and the appropriate solution is the one where the smallest $|n|$ leads to the correct signs for $\alpha_1$ and $\beta_1$. In order to make calculations more convenient, Ref.~\cite{Zhong:2024qqs} has proposed the new formula
\begin{align} \label{newrelation}
\delta_P-\delta_S & \,=\, 2 \arctan\frac{\beta}{\sqrt{\alpha^2+\beta^2}+\alpha} \,. ~~~
\end{align}
It yields a reliable determination of the phase shift in Eq.\,(\ref{DefineIdirive}) and naturally derive the correct solution, which belongs to $[-\pi,\pi]$.

The importance of the phase shift can be seen in the $\alpha$-based $CP$-violating quantity defined as \,$A^{\alpha}_{CP} = \big(\alpha+\overline\alpha\big)/\big(\alpha-\overline\alpha\big)$\,~\cite{Donoghue:1986hh}, with $\overline\alpha$ belonging to the antiparticle mode, especially if the {\tt S}- and {\tt P}-wave amplitudes each contain only one term (or are dominated by one). 
In that case, if $CP$ violation is present, the amplitudes in Eq.\,(\ref{DefineI}) turn (approximately) into \,$S_1=|S|e^{i(\delta_S+\xi_S)}$ and \,$P_1=|P|e^{i(\delta_P+\xi_P)}$,\, their antiparticle counterparts are given by \,$\overline S_1=-|S|e^{i(\delta_S-\xi_S)}$\, and \,$\overline P_1=|P|e^{i(\delta_P-\xi_P)}$,\, with $\xi_{S,P}$ denoting weak-interaction phases,\, and consequently 
\begin{equation} \label{CPV_phaseshift}
A^\alpha_{CP} \,=\, {-}\tan(\delta_P-\delta_S) \tan(\xi_P-\xi_S) \,.
\end{equation}
Examples of processes having these properties are the hyperon modes $\Lambda\to p\pi^-$ and $\Xi^-\to\Lambda\pi^-$ \cite{Donoghue:1986hh} and the Cabibbo-favored decay $\Lambda_c^+
\to\Xi^0K^+$ \cite{Geng:2023pkr,Zhong:2024qqs}. 
Since the standard model makes specific predictions for $A^\alpha_{CP}$, precise measurements of it in such decays would test the model stringently.\footnote{More generally the $\tt S$ and $\tt P$ waves may each be a sum of two or more terms, with no dominant one, causing $A^\alpha_{CP}$ to be more complicated than that in Eq.\,(\ref{CPV_phaseshift}) and consist of a number of terms \cite{Donoghue:1986hh}. 
The latter might have different signs and thus yield conflicting effects on $A^\alpha_{CP}$, which could suppress its size.}

Recently, the phase shift in $\Lambda_c^+\to\Xi^0K^+$ has been measured for the first time~\cite{BESIII:2023wrw}, and the reported value is significant, $-1.55\pm0.25~\text{rad}$ or $1.59\pm0.25~\text{rad}$, 
the two possibilities being due to substantial statistical uncertainty~\cite{BESIII:2023wrw}. 
The sizable phase shift is an enhancing factor for $A^{\alpha}_{CP}$ in this decay and has motivated further theoretical work~\cite{Zhong:2024qqs,Geng:2023pkr}.

Besides the expressions in Eq.~\eqref{DefineI}, other conventions can be found in various experimental analyses and theoretical computations. In particular, if $CP$ is again conserved, the following second and third parameterization forms have been adopted:
\begin{align} \label{DefineII}
S_2 &= \widetilde{S} e^{i\delta_S}, & P_2 &= \widetilde{P}\, e^{i\delta_P} \,, \\ \label{DefineIII}
S_3 &= |\mathcal{A}| \sin\zeta\, e^{i\delta_S}, & P_3 &= |\mathcal{A}| \cos\zeta\, e^{i\delta_P} \,.
\end{align} 
In the second one \cite{Donoghue:1986hh,Cronin:1963zb,Overseth:1967zz}, Eq.\,(\ref{DefineII}), $\widetilde{S}$ and $\widetilde{P}$ are real constants, one or both of which may be negative. In the third convention~\cite{Salone:2022lpt}, Eq.\,(\ref{DefineIII}), $\zeta$ is a tuning parameter that ranges from $0$ to $\pi$, and consequently \,$|\mathcal{A}|^2=|S_3|^2+|P_3|^2$.\, 
The corresponding polarization parameters are then derived to be
\begin{align} \label{DefineIIderive} 
\alpha_2 &= \frac{2\widetilde{S}\widetilde{P} \cos(\delta_P-\delta_S)}{|\widetilde{S}|^{2}+|\widetilde{P}|^{2}}, & \beta_2 &= \frac{2\widetilde{S}\widetilde{P} \sin(\delta_P-\delta_S)}{|\widetilde{S}|^{2}+|\widetilde{P}|^{2}} \,, 
\\ \label{DefineIIIderive}
\alpha_3 &= \sin(2\zeta) \cos(\delta_P-\delta_S), & \beta_3 &= \sin(2\zeta) \sin(\delta_P-\delta_S) \,. 
\end{align}

If, for example, $\alpha$ is positive, \,$\delta_P-\delta_S$\, from Eqs.~(\ref{DefineIIderive})-(\ref{DefineIIIderive}) becomes \,$\arctan(\beta_{2,3}/\alpha_{2,3})$\, when $\widetilde{S}\widetilde{P}$ and $\text{sin}(2\zeta)$, respectively, are positive and \,$\arctan(\beta_{2,3}/\alpha_{2,3})\,{+}j\pi$\, ($j$\,=1\,or~$-1$) when either is negative, as the arctan function returns a value only within the interval \,[$-\pi/2$, $\pi/2$].
Thus, when using the convention in Eq.~\eqref{DefineII} or \eqref{DefineIII}, the solution for $\delta_P-\delta_S$ is sensitive to the sign of $\widetilde{S}\widetilde{P}$ or $\sin(2\zeta)$. 
However, this sign is not always known or explicitly stated.

Here we are interested in having a convenient way for evaluating the phase shift which can accommodate the different parameterization options. 
To that end, we introduce 
\begin{equation} \label{newrelation2}
\delta_P-\delta_S \,=\, 2\arctan\frac{\beta\times sign}{\sqrt{\alpha^2+\beta^2}+\alpha\times sign} \,, ~~~
\end{equation}
which is a modified version of Eq.\,(\ref{newrelation}) and incorporates a constant ``$sign$'' which may have a value of +1 or $-1$.
Clearly, the expression in Eq.\,(\ref{newrelation2}) will revert to that in Eq.\,(\ref{newrelation}) in the first convention if $sign$ is equated to +1. 
Therefore, Eq.\,(\ref{newrelation2}) may be regarded as a unified parameterization form, which is applicable to the three conventions discussed here.

The $\alpha$, $\beta$, and $\gamma$ parameters can also be described within the helicity framework \cite{Lee:1957qs,ParticleDataGroup:2024cfk}. One angle $\phi$ is introduced, which means the phase difference between two different helicity amplitudes, and then $\beta$ and $\gamma$ can be derived from the relations $\beta =\sqrt{1-\alpha^2} \sin\phi$\, and \,$\gamma = \sqrt{1-\alpha^2} \cos\phi$.

\begin{table*}[!t] \centering \footnotesize
\caption{\small Summary of experimental data on the parameters $\alpha$, $\beta$, and $\phi$ and the phase shift, $\delta_P-\delta_S$, in various two-body nonleptonic decays of the $\Lambda$ and $\Xi$ hyperons and singly charmed baryon $\Lambda_c^+$.
} \label{summary} \setlength\tabcolsep{0.7ex} \footnotesize

    \begin{tabular}{ccccccccc}
		\hline
            \multirow{2}*{Experiment}                                 & \multirow{2}*{Process}                       & \multirow{2}*{$\alpha$ or $\langle\alpha\rangle$} & \multirow{2}*{$\beta$ or $\langle\beta\rangle$} & $\phi$ or $\langle\phi\rangle$ & $\delta_P-\delta_S$              & Value   & \multicolumn{2}{c}{$\delta_P-\delta_S$ (rad)}  \\
            ~                                                         & ~                                            & ~                                                 & ~                                               & (rad)                          & (rad)                            & of $sign$    & \eqref{newrelation2} with $sign$\,=1 & \eqref{newrelation2} with $sign$\,=$-1$ \\ 
            \hline
            \!$\Lambda$ from $\pi^-p$ (1963) \cite{Cronin:1963zb}\!   & \multirow{2}*{$\Lambda\to p\pi^-$}           & $0.62\pm0.07$                                     & $-0.18\pm0.24$                                  & \raisebox{0.5ex}{...}          & \!\!$-0.26\pm0.35\,^\dagger$\!\! & Unknown                & $-0.28\pm0.36$       & $2.86\pm0.36$ \\
            \!$\Lambda$ from $\pi^-p$ (1967) \cite{Overseth:1967zz}\! & ~                                            & $0.645\pm0.017$                                   & $-0.103\pm0.065$                                & $-0.14\pm0.10$                 & \!\!$-0.16\pm0.10\,^\dagger$\!\! & Unknown                & $-0.16\pm0.10$       & $2.98\pm0.10$ \\
            E756 (2003) \cite{FNALE756:2003kkj}                       & \multirow{5}*{$\Xi^-\to\Lambda\pi^-$}        & $-0.458\pm0.012$                                  & $-0.03\pm0.04$                                  & $-0.03\pm0.05$                 & ~ $0.06\pm0.09$                  & $+1$               & $-3.08\pm0.09$       & $ 0.06\pm0.09$ \\
            HyperCP (2004) \cite{HyperCP:2004not}                     & ~                                            & $-0.458\pm0.012$                                  & $-0.037\pm0.015$                                & $-0.041\pm0.016$               & ~ $0.080\pm0.032$                & Unknown                & $-3.062\pm0.031$     & $ 0.079\pm0.031$ \\
            BESIII~(2022)~\cite{BESIII:2021ypr}                       & ~                                            & $-0.373\pm0.006$                                  & positive                                        & ~ $0.016\pm0.016$              & $-0.040\pm0.037$                 & Unknown                & $3.102\pm0.036$      & $-0.040\pm0.036$ \\
            BESIII~(2022)~\cite{BESIII:2022lsz}                       & ~                                            & $-0.350\pm0.018$                                  & positive                                        & ~ $0.073\pm0.052$              & $-0.20\pm0.13$                   & Unknown                & $ 2.95\pm0.13$       & $-0.19\pm0.13$ \\
            BESIII~(2024)~\cite{BESIII:2023jhj}                       & ~                                            & $-0.371\pm0.004$                                  & negative                                        & $-0.013\pm0.008$               & ~ $0.033\pm0.023$                & Unknown                & $-3.109\pm0.025$     & $0.033\pm0.025$ \\
            BESIII~(2023)~\cite{BESIII:2023drj}                       & $\Xi^0\to\Lambda\pi^0$                       & $-0.377\pm0.003$                                  & positive                                        & $0.005\pm0.007$                & $-0.013\pm0.017$                 & Unknown                & $3.129\pm0.017$      & $-0.012\pm0.017$  \\
            LHCb~(2024)~\cite{LHCb:2024tnq}                           & $\Lambda_c^+\to\Lambda\pi^+$                  & $-0.785\pm0.007$                                  & $0.378\pm0.015$                                 & $0.656\pm0.027$                & $2.693\pm0.017$                 & Unknown                & $2.693\pm0.015$       & $-0.449\pm0.015$\\            
            LHCb~(2024)~\cite{LHCb:2024tnq}                           & $\Lambda_c^+\to\Lambda K^+$                  & $-0.516\pm0.046$                                  & ~ $0.33\pm0.08$                                 & $2.75\pm0.11$                  & $2.57\pm0.19$                   & Unknown                & $2.58\pm0.12$         & $-0.56\pm0.12$\\
            BESIII~(2024)~\cite{BESIII:2023wrw}                       & \!$\Lambda_c^+\to\Xi^0 K^+$\!                & ~ $0.01\pm0.16$                                   & $-0.64\pm0.70$                           & $3.84\pm0.90$                  & \!$-1.55(1.59)\pm0.25^\dagger$\!         & $+1$               & $-1.55(1.59)\pm0.25$  & $1.59(-1.55)\pm0.25$\\
            \hline
            \multicolumn{7}{l}{\footnotesize$^\dagger$\,The value inferred from \,$\pi^-p\to\pi^-p$\, scattering data is \,$(-0.11\pm0.03)$\,rad\, \cite{Overseth:1967zz,Barnes:1960zz}.}\\
            \multicolumn{7}{l}{\footnotesize$^\ddagger$\,The phase shift has two solutions because of the large uncertainty of $\alpha$.}
	\end{tabular}
\end{table*}

In Table~\ref{summary} we have collected the experimental data on the parameters $\alpha$, $\beta$, and $\phi$ and the phase shift in several two-body nonleptonic decays of the $\Lambda$ and $\Xi$ hyperons and of the singly charmed baryon $\Lambda_c^+$, including the aforementioned BESIII finding of large $|\delta_P-\delta_S|$ in $\Lambda_c^+\to\Xi^0 K^+$ \cite{BESIII:2023wrw}. 
In the columns fourth to sixth, the values are quoted from the original published text, and some of them are averages ($\langle\alpha\rangle$, $\langle\beta\rangle$, and $\langle\phi\rangle$).
In the seventh column, we give the values of $sign$ if known from the papers cited in the table. 
For comparison of alternative schemes, the last two columns contain the phase shifts obtained using Eq.~\eqref{newrelation2} with $sign=+1$ and $-1$, respectively, where slight differences between the results and those reported in the original papers are due to differences in computational accuracy.
Evidently, the evaluated \,$\delta_P-\delta_S$\, differ by $\pi$ between opposite $sign$ choices, but such a difference has no effect on the value of $\tan(\delta_P-\delta_S)$.

From the table, it is important to note additionally that in the experimental analysis of Ref.\,\cite{FNALE756:2003kkj} on $\Xi^-\to\Lambda\pi^-$  the amplitudes were defined according to Eq.\,\eqref{DefineI}, implying  $sign=+1$,
and  $\delta_P-\delta_S=0.06\pm0.09$\,  was reported, implying \,$\alpha>0$\, should have been found, but this was not the case. 
The reason for this lack of consistency, also evident in the table, is that the relation between $\alpha$, $\beta$, and \,$\delta_P-\delta_S$\, was not well considered when solving the arctan function, which has a range of \,[$-\pi/2$, $\pi/2$]. 

The table further shows that for the same $sign$ choice the phase shifts in the $\Lambda$ and $\Xi$ modes are distinctly different, with one approaching 0 and the other near $\pi$. 
However, the \,$\Lambda\to p\pi^-$  phase shift can already be independently cross-checked with  $p\pi^-\to p\pi^-$\, scattering data~\cite{Salone:2022lpt,Overseth:1967zz,Barnes:1960zz}, unlike \,$\Xi\to\Lambda\pi$.

Lastly, the table also includes the aforesaid $\Lambda_c^+ \to \Xi^0 K^+$ results as well as the LHCb measurements of \,$\Lambda_c^+ \to \Lambda \pi^+$\, and \,$\Lambda_c^+ \to \Lambda K^+$~\cite{LHCb:2024tnq}, confirming with high precision the presence of nonzero phase shifts in $\Lambda_c^+$ decays.
These findings provide valuable insights into the underlying dynamics of baryons.

To conclude, the strong phase shift is an important ingredient in the study of $CP$ violation among baryons~\cite{Donoghue:1986hh,Zhong:2024qqs,He:2024pxh,Geng:2023pkr}, in order to pin down its sources within the standard model and possibly beyond it, and a sizable phase shift might greatly enhance the magnitude of certain $CP$ asymmetries.
Many experiments have reported measurements of the strong phase shifts in various two-body nonleptonic baryon decays.
However, different experimental and theoretical studies have employed different conventions for the weak decay amplitudes, and the resulting solutions for the strong phase shifts suffer from ambiguities in some instances. 
In this paper, we have presented a brief review of the situation and proposed a unified parameterization form in Eq.~\eqref{newrelation2} to accommodate the different conventions.
To avoid potential ambiguities, we recommend that future studies provide clear parameterization methods, particularly for $sign$, to enable discussions based on a unified definition.
Although, as remarked above, the $sign$ choice does not affect the tangents of the phase shifts, 
awareness of this issue is important when conducting global analyses of baryon processes that involve the strong phase shifts obtained from different experiments and performing future searches for baryon $CP$ violation.

\section*{Conflict of interest}
The authors declare that they have no conflict of interest.

\section*{Acknowledgments}
The authors thank Fu-Sheng Yu and Fan-Rong Xu for useful discussions. J.T. thanks the Institute of High Energy Physics (IHEP), Chinese Academy of Sciences (CAS), for kind hospitality and support during the early stage of this research.
This work is supported in part by National Key R\&D Program of China under Contract Nos. 2023YFA1609400, 2020YFA0406400, 2023YFA1606000; National Natural Science Foundation of China (NSFC) under Contracts Nos. 11935018, 12221005, 12422504, 12105127, 124B2097;
Fundamental Research Funds for the Central Universities, Lanzhou University under Contracts Nos. lzujbky-2021-ey09, lzujbky-2023-it32. 

%

\end{document}